# Application of Gaussian Processes to online approximation of compressor maps for load-sharing in a compressor station

Akhil Ahmed, Marta Zagorowska, Ehecatl Antonio del Rio-Chanona and Mehmet Mercangöz*

*Abstract* — Devising optimal operating strategies for a compressor station relies on the knowledge of compressor characteristics. As the compressor characteristics change with time and use, it is necessary to provide accurate models of the characteristics that can be used in optimization of the operating strategy. This paper proposes a new algorithm for online learning of the characteristics of the compressors using Gaussian Processes. The performance of the new approximation is shown in a case study with three compressors. The case study shows that Gaussian Processes accurately capture the characteristics of compressors even if no knowledge about the characteristics is initially available. The results show that the flexible nature of Gaussian Processes allows them to adapt to the data online making them amenable for use in real-time optimization problems.

## I. INTRODUCTION

The transport of natural gas along pipelines can range over several thousand kilometers and result in pressure losses due to friction. These pressure losses are compensated by compressor stations at regular intervals such that the pressure of the gas is boosted to ensure adequate transport to the destination of choice. Compressor stations consist of compressors arranged in a parallel or series formation which is typically decided based on a techno-economic analysis [1]. These compressors will have performance characteristics defined in the form of compressor maps, which express the efficiency of the compressor as a function of mass flow and pressure ratio. These performance characteristics tend to differ from one machine to the other and will vary over time due to degradation phenomena [2]. For compressors with variable speed drivers, the compressor operating point can be manipulated to alter the compressor efficiency to minimize the power consumption. However, when multiple compressors are combined as part of a compressor station, all with different performance characteristics, the minimization of the power consumption of the whole station becomes a complex optimization problem referred to as load sharing optimization (LSO) [3] [4] [5] [6] [7]. For parallel formations, which will be the focus of this paper, load sharing optimization attempts to solve the problem of mass flowrate allocation to each of the compressors. Effectively, the compressor station will get a mass flowrate target for the whole station from a dispatch center. Load sharing optimization then attempts to determine the allocation of the total flow to each of the individual compressors which minimizes the total power consumption of the station. The solution to this problem, which are the individual compressor mass flowrate targets, are then sent as set points to low-level controllers for each of the machines which attempts to track this reference. This is actuated by adjusting the torque applied to the compressors and hence the compressor speed.

The complexity of the problem arises from the fact that an accurate model of the compressor performance characteristics is necessary to achieve an accurate solution to the optimization problem. Without an up-to-date and accurate model, there is a risk that the compressor station can be operated in a sub-optimal manner due to plant-model mismatch. This is a known issue in the real-time optimization literature and several adaptation strategies have been proposed to this end [8]. Namely [8] classifies three primary approaches; (i) *Model-parameter adaptation* or the *two-step approach* whereby the model parameters are estimated and updated based on output measurements from the system before the updated model is used for optimization. (ii) *Modifier adaptation* where the cost function and constraints are modified before optimization is performed to ensure the model and plant share the same optimality conditions. (iii) *Direct input adaptation* whereby the optimization problem is reformulated into a feedback control problem. Specifically, *model-parameter adaptation* and *modifier adaptation* have been previously used to address the issue of load sharing optimization applied to compressor stations. In particular, both [6] and [7] use the *modifier adaptation* approach to ensure the model shares the same optimality conditions as the plant. However, as explained in both [6] and [8], a major challenge for the *modifier adaptation* approach is the accurate estimation of the gradients of the cost function and/or constraints. In both [6] and [7], this issue is alleviated by exploitation of the problem structure which allows for the gradients to be estimated, however, this may not generally be the case. In addition to this, *modifier adaptation* can be highly sensitive to noisy measurements which may make the approach less robust when the level of measurement noise is significant as discussed in [8] and [9].

On the other hand, [3], [4] and [5] adopted the *model-parameter adaptation* approach in order to overcome the issue of plant-model mismatch. However, as explained in [8], the robustness of this approach relies on the ability of the adaptation scheme to reduce the plant-model mismatch. Where a particular parametric model is used to represent the system of interest, a structural mismatch may exist between the model and plant if the model structure is not sufficient to capture the real behavior of the plant. The challenge with compressor performance maps is that a structurally correct representation is seldom available as degradation of

A. Ahmed, M. Zagorowska, E.A. del Rio-Chanona and M. Mercangöz (*corresponding author) are with the Department of Chemical Engineering, Imperial College London. (e-mail: {a.ahmed21, m.zagorowska, a.del-rio-chanona, m.mercangoz}@imperial.ac.uk; phone: +44 (0)20 7594 1191).

compressor performance typically occurs over the lifetime of the compressor. As a result, a parametric model of the compressor map, such as a polynomial expression as used in [3], [4] and [5], may quickly begin to exhibit a structural mismatch after some time.

However as discussed in [9], if the model used is flexible enough to overcome any structural plant-model mismatch then the authors of [9] argue that the classical method of *model-parameter adaptation* is a robust and reliable approach when considering different degrees of plant-model mismatch as well as measurement noise. For this reason, in this paper we propose the use of a non-parametric model, namely Gaussian Processes (GP) [10]. The chief advantage of using a GP is that, as a non-parametric model, no underlying assumptions regarding the functional form of the system to be represented are made. Consequently, GPs are highly flexible, and the issue of structural mismatch can be avoided, and the overall plant-model mismatch can be minimized. This addresses the major limitations of the *model-parameter adaptation* approach as discussed in [8] and [9]. As demonstrated in the rest of the paper, the use of Gaussian Processes in the *model-parameter adaptation* approach to the load sharing optimization problem is a robust technique which can handle varying degrees of plant-model mismatch and adapt accordingly to uncertainty.

## II. THE LOAD SHARING OPTIMIZATION PROBLEM

### A. Problem formulation

In this paper, we will consider the problem of load sharing optimization applied to the parallel arrangement of three compressors as depicted in Fig 1. In this arrangement, load sharing optimization attempts to distribute the station mass flow rate target to each of the machines to minimize the total power consumption of the station. The objective function can be written as:

$$J(\dot{m}_i) = \sum_{i=1}^{N_c} P_i(\dot{m}_i, \eta_i) \quad (1)$$

where $N_c$ is the number of compressors (for this case $N_c = 3$), $i \in \mathbb{Z}^+$ is the compressor index, $\dot{m}_i \in \mathbb{R}$ is the $i^{th}$ compressor mass flowrate, $P_i \in \mathbb{R}$ is the power consumption of the $i^{th}$ compressor and $\eta_i \in \mathbb{R}$ is the compressor efficiency of the $i^{th}$ compressor. As explained in section I, the compressor map expresses the compressor efficiency, $\eta_i$, as a function of both the compressor mass flowrate and pressure ratio of the compressor which we denote as $\Pi_i \in \mathbb{R}$. The power consumption of the compressor can be expressed as:

$$P_i = \frac{y_{p,i}}{\eta_i} \dot{m}_i \quad (2)$$

where $y_{p,i} \in \mathbb{R}$ is the polytropic head of compressor i and is defined as:

$$y_{p,i} = \frac{Z_{in} R T_{in}}{MW} \frac{n_v}{n_v - 1} \left[ \Pi_i^{\left(\frac{n_v}{n_v-1}\right)} - 1 \right] \quad (3)$$

where R is the universal gas constant, $n_v \in \mathbb{R}$ is the polytropic exponent, $MW \in \mathbb{R}$ is the molecular weight of the compressed gas mixture, $Z_{in} \in \mathbb{R}$ is the inlet compressibility factor while $T_{in} \in \mathbb{R}$ denotes the inlet compressor temperature. Consequently, the load sharing optimization problem for the parallel arrangement can be formulated as follows:

$$\min_{\{\dot{m}_1, \ldots, \dot{m}_{N_c}\}} J(\dot{m}_i) \quad (4a)$$

$$\text{s.t. } \mathbf{h}(\dot{m}_1, \ldots, \dot{m}_{N_c}) = \mathbf{0} \quad (4b)$$

$$\mathbf{g}(\dot{m}_1, \ldots, \dot{m}_{N_c}) \leq \mathbf{0} \quad (4c)$$

$$\sum_{i=1}^{N_c} \dot{m}_i = \dot{m}_{station} \quad (4d)$$

where $\mathbf{h} : \mathbb{R}^{N_c} \to \mathbb{R}^{N_h}$ (4b) defines the equality constraints imposed by the steady state model equations of the compressor. Similarly, $\mathbf{g} : \mathbb{R}^{N_c} \to \mathbb{R}^{N_g}$ (4c) defines the surge and choke inequality constraints which are formulated to prevent unstable compressor operation due to surge or choke conditions [5]. Both (4b) and (4c) are further discussed in section III and are based on [7]. Finally, (4d) defines an additional equality constraint that the sum of the compressor mass flowrates must equal the station mass flow rate target, $\dot{m}_{station}$.

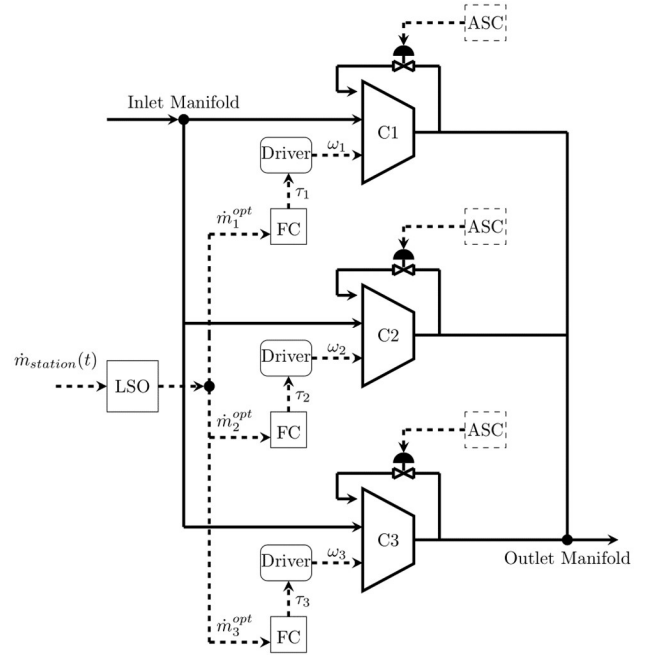

Figure 1. Schematic diagram of the LSO problem applied to three compressors arranged in parallel. FC represents the compressor flow controllers which actuate a change in the torque, $\tau_i$, applied to the compressors which adjusts the operating speed of the compressors, $\omega_i$. ASC denotes the Anti-Surge Controllers which are designed to recycle gas flow to the compressors to prevent surging. Control signals are represented by dashed lines and gas flows are represented by solid lines.

### B. Problem formulation for the model-parameter adaptation approach using Gaussian Processes

As mentioned in section I, an accurate solution to the optimization problem is contingent on an accurate model of the compressor performance characteristics:

$$\hat{\eta}_i = \hat{f}(\dot{m}_i, \Pi_i) \quad (5)$$

where $\hat{f} : \mathbb{R}^2 \to \mathbb{R}$ defines the model of the compressor map while $\hat{\eta}_i$ defines the model compressor efficiency. Similarly, we express the real compressor performance characteristics as:

$$\eta_i = f(\dot{m}_i, \Pi_i) \qquad (6)$$

where $f : \mathbb{R}^2 \to \mathbb{R}$ represents the true function relating the compressor mass flowrate and pressure ratio to the efficiency of the compressor. The goal of the *model-parameter adaptation* approach is to minimize the difference between the true function, f, and the approximation, $\hat{f}$, as the system is operated and more information about the true behaviour of the system is obtained.

For the reasons outlined in section I, in this paper we propose the use of Gaussian Processes to learn the compressor performance characteristics. However, it should be noted that for our adaptation approach two primary alternatives are considered. In the first and main focus of investigation for this paper, the GPs are not used to directly learn the compressor performance characteristics. Instead, similar to [3], [4] and [5], a second order polynomial is used as an initial model for the performance characteristics, which we refer to as a prior model:

$$\hat{\eta}_i^{poly} = \alpha_1 + \alpha_2 \dot{m}_i + \alpha_3 \Pi_i + \alpha_4 \dot{m}_i \Pi_i + \alpha_5 \dot{m}_i^2 + \alpha_6 \Pi_i^2 \qquad (7)$$

where $\alpha_j \in \mathbb{R}$ are the polynomial coefficients and $\hat{\eta}_i^{poly} \in \mathbb{R}$ is the polynomial efficiency.

As the system is operated and more data is obtained from the compressors the difference between the predicted efficiency and estimated plant efficiency (which is back calculated from the plant measurements using (2) and (3)) up to the $k^{th}$ sampling instant can be determined:

$$\Delta_i^k = \eta_i^k - \hat{\eta}_i^{poly,k} \qquad (8)$$

where $\eta_i^k \in \mathbb{R}^k$ and $\hat{\eta}_i^{poly,k} \in \mathbb{R}^k$ represent the vectors of estimated plant efficiency and efficiency predicted by the polynomial up to the $k^{th}$ sampling instant respectively while $\Delta_i^k \in \mathbb{R}^k$ represents the vector of their difference. Consequently, an error function, $\Delta(\dot{m}_i, \Pi_i) : \mathbb{R}^2 \to \mathbb{R}$, represented by a GP can be fitted on the input ($\dot{m}_i^k, \Pi_i^k$) and output dataset ($\Delta_i^k$) where the GP can be thought of as a multivariate Gaussian distribution over functions from which a function, $g : \mathbb{R}^2 \to \mathbb{R}$, can be sampled:

$$g(\dot{m}_i, \Pi_i) \sim GP(m(\cdot), k(\cdot, \cdot)) \qquad (9)$$

$$\Delta(\dot{m}_i, \Pi_i) = m(\cdot) \qquad (10)$$

where $m(\cdot)$ and $k(\cdot, \cdot)$ represent the mean function and covariance function of the GP trained with the input-output dataset.

Finally, the model efficiency is expressed as the sum of the prior model and the fitted error function:

$$\hat{\eta}_i(\dot{m}_i, \Pi_i) = \hat{\eta}_i^{poly}(\dot{m}_i, \Pi_i) + \Delta(\dot{m}_i, \Pi_i) \qquad (11)$$

As a result, with this approach, the model should approximate the plant behaviour if the GP learns the error function adequately. The utility of this approach is two-fold. Firstly, this approach provides the ability to incorporate prior information into the adaptation scheme. Namely, if there exists a model of the system in which there is some confidence, then this can be used as a prior for the GP i.e. as an initial starting point. Secondly, the advantage of using the GP as an error function lies in the fact that, as a non-parametric model, the GP can overcome any structural mismatch between the prior parametric model (polynomial in this case) and the real plant. This would not be possible if a *model-parameter adaptation* approach was used directly on the prior parametric model as a structural mismatch would remain if the model structure was inadequate. This is indeed the case when a second order polynomial is used to represent the complex structure of the compressor map.

The second focus of investigation uses GPs to directly learn the compressor map. That is, no prior model is assumed for the compressor map and instead this is learned directly by the GPs from data measured during operation relating the input ($\dot{m}_i^k, \Pi_i^k$) to the output dataset ($\eta_i^k$). This can be expressed as:

$$h(\dot{m}_i, \Pi_i) \sim GP(m(\cdot), k(\cdot, \cdot)) \qquad (12)$$

$$\hat{\eta}_i(\dot{m}_i, \Pi_i) = m(\cdot) \qquad (13)$$

By having these two different lines of investigation, it is possible to examine the effect of different degrees of mismatch between the prior model and plant and especially the ability of the GPs to overcome this mismatch. Additionally, this also allows us to compare the difference in performance between a hybrid approach, where a prior model is used, and a purely data-driven approach where instead a model is learned directly from data.

## III. SIMULATION SET-UP

Four main sets of cases were simulated in this investigation for comparison purposes. Before describing each of these cases and the relationship between them, we first provide a description of the components which make up the general simulation although the exact structure of the simulation will vary for each of the cases. Any differences will be explained upon discussion of the cases considered.

The general simulation structure is summarized in Fig. 2.

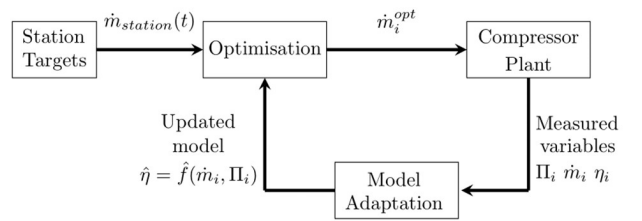

Figure 2. Block diagram of the simulation set-up used in this study where the station target, $\dot{m}_{station}$, is sent to the optimization block as well as an updated compressor model from the model adaptation block so that the LSO problem can be solved. After implementing the calculated set-points, measurements from the system are used to update the model in the adaptation block if necessary. Station targets change regularly over time.

There are four main components to the simulation which are described in the following sections. The simulations were performed to mimic a period of three days of operation with new station targets being sent at regular intervals.

### A. Station target block

The station target block defines the compressor station mass flowrate targets over the course of three days of operation as depicted in Fig. 3. The mass flowrate profiles are defined to mimic expected daily gas consumption profiles. This includes a ramp up in the station target up to a peak load and an eventual ramp down [11] [12]. The station target is sent to the optimization block so that the load sharing optimization problem can be solved, and the target mass flowrate distributed to the compressors accordingly.

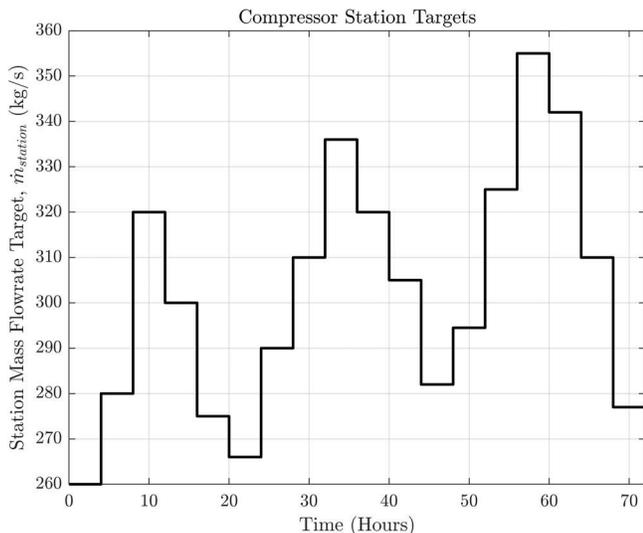

Figure 3: Mass flowrate targets for the compressor station defined for three days of simulation.

### B. Optimization block

The optimization block solves the load sharing optimization problem presented in Section II. It does this by taking the station mass flowrate target $\dot{m}_{station}$ to impose the constraint defined in (4d) and solving the nonlinear optimization problem formulated in (4) with the model of the compressor map, $\hat{\eta}_i=\hat{f}(\dot{m}_i,\Pi_i)$ defined using GPs. The solution of the optimization problem, consisting of the individual compressor mass flowrate targets, is then sent as set points to the corresponding controllers as shown in Fig. 1. To avoid frequent changes of the set points sent to the compressors, the optimization is only performed when the station target reference entering the optimization block changes.

As discussed in Section I, the model compressor map $\hat{\eta}_i=\hat{f}(\dot{m}_i,\Pi_i)$ and the plant compressor map $\eta_i = f(\dot{m}_i,\Pi_i)$ should exhibit no mismatch in order to ensure the optimum obtained from the optimization block corresponds to the true optimum of the plant.

### C. Model adaptation block

At set sampling instants, the model adaptation block takes the measurements of mass flowrate, pressure ratio, power, and temperature from the compressors to estimate the real plant efficiency using (2) and (3). Each of these measurements are corrupted with measurement noise at a level similar to that of [13]. As discussed in [3] and [8], it is important to ensure that model adaptation is only triggered when new data is measured i.e. different to what has already been observed. This is especially relevant for GPs which exhibit performance issues with larger datasets [10].

For this reason, in the first instance, a Euclidean distance metric is used in the model adaptation block to determine whether the data measured is similar to any other data in the current dataset used to fit the GP. If it is, then the data is discarded, and the adaptation method is terminated. Otherwise, the data is appended to the current dataset and the GP is re-fitted.

### D. Compressor block

The compressor block consists of a number of components which can be inferred from Fig. 1 and are described in this section for clarity.

Firstly, the compressor block contains the *compressor model* which consists of the full set of differential and algebraic equations which describes the dynamic and steady-state behaviour of the compressor. These equations have been adapted from [7]. Most importantly, the *compressor model* contains the plant compressor map i.e. $\eta_i = f(\dot{m}_i,\Pi_i)$. This information is not available to any other components of the simulation and represents the baseline of the compressor map. Fig. 4 depicts the compressor maps for compressor 1, 2 and 3 respectively. It should be noted that for the purposes of the simulation, the efficiencies were scaled such that compressor 1 was the most efficient compressor, followed by compressor 2 and 3.

Secondly, a low-level *PI flow controller* receives the individual compressor mass flowrate target from the optimization block as a set-point. This set-point is tracked by adjusting the torque applied to the compressor and hence adjusting the compressor operating speed.

Finally, an *anti-surge controller* acts to ensure stable compressor operation by preventing the compressor from operating in the surge region [5]. However, the surge and choke constraints defined in (4c) have been formulated specifically to ensure that surging or choking of the compressor is avoided without needing the action of the *anti-surge controller*.

The main components of the simulation described in this section have been used in four primary case studies discussed in section IV.

## IV. CASES SIMULATED

The four main sets of cases simulated in this investigation are summarized in this section. All the cases were run in MATLAB/Simulink and the optimization problem was solved using fmincon.

### A. Case 1

Case 1 simulates the scenario where no plant-model mismatch exists. That is, in the optimization block described in section III.B, the model of the compressor map is identical to the plant compressor map defined in section III.D:

$$\hat{\eta}_i = \eta_i = f(\dot{m}_i, \Pi_i) \quad (13)$$

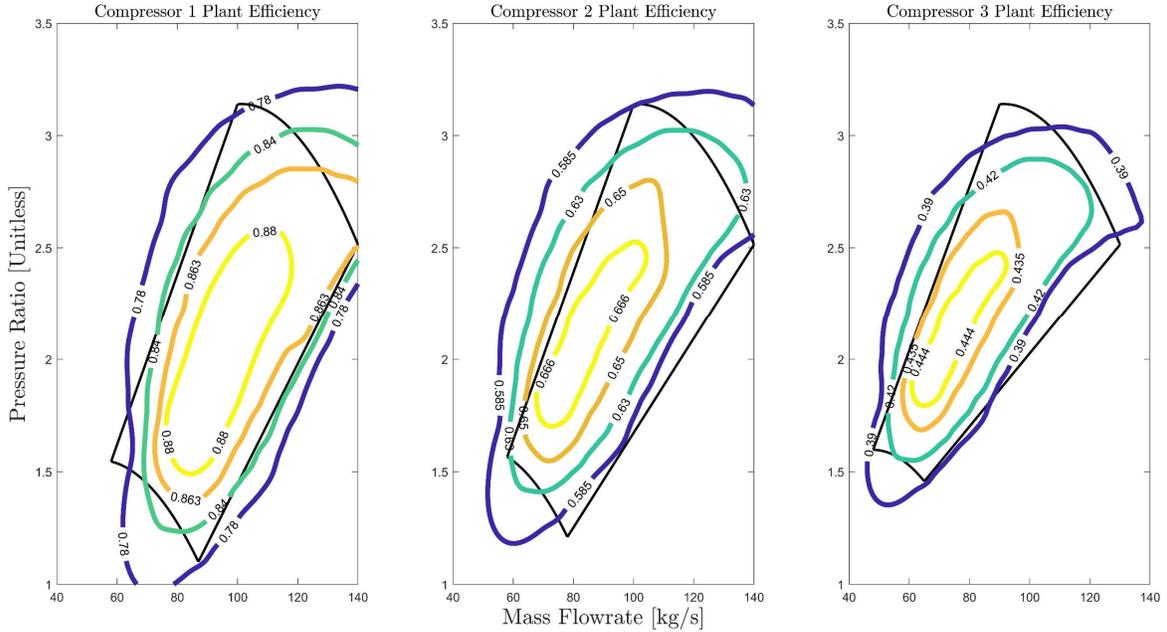

Figure 4. Baseline compressor maps for compressors 1,2 and 3. The efficiencies were scaled to ensure variation amongst the compressors with compressor 1 being the most efficient followed by compressor 2 and 3. Black solid line represents the operating envelope of the compressors.

The purpose of this case is to bound the problem being studied by giving a lower bound on the expected power consumption i.e. a best case scenario. Consequently, this case requires no model adaptation block as a result.

### B. Case 2.1-2.3

Cases 2.1-2.3 simulate the scenario where a plant-model mismatch exists, and this mismatch is not addressed i.e. no model parameter adaptation is performed. This is done by using second order polynomials of the form of (7) for the model of the compressor map in the optimization block. Furthermore, several sub-cases are defined where the second order polynomials are fitted on limited amounts of real data to induce varying degrees of plant-model mismatch. Specifically, 2, 5 and 20 points are used in the polynomial regression which we define as case 2.1, 2.2 and 2.3 respectively. Additionally, a further mismatch is induced by shifting the efficiency surfaces produced by the resulting second order polynomials by a random number. An example of this mismatch is demonstrated for compressor 3 for case 2.1 in Fig. 5. As can be seen, a mismatch exists between the plant and model due to the use of only 2 points for the polynomial regression as well as the applied shift.

In the same vein as case 1, cases 2.1-2.3 are intended to bound the problem by giving an upper bound on the expected power consumption i.e. a set of worst case scenarios. Additionally, as will be discussed next, by using different amounts of data for the polynomial regression, this allows us to investigate the effect of varying degrees of plant-model mismatch when the GP error functions are used for cases 3.1-3.3. In particular, this allows us to test the robustness of the GPs to varying degrees of model uncertainty.

### C. Case 3.1-3.3

Cases 3.1-3.3 define the main focus of investigation of this paper. This is the scenario described in section II.B, where the second order polynomials of case 2.1, 2.2 and 2.3 are used as prior models for GPs which attempt to learn the error between the prior model and the plant. Consequently, 3 sub-cases are defined: case 3.1, 3.2 and 3.3 which have the second order polynomials fitted on 2, 5 and 20 data points as prior models respectively. The model of the compressor map used by the optimization block is (11), the sum of the GP error function and the prior model.

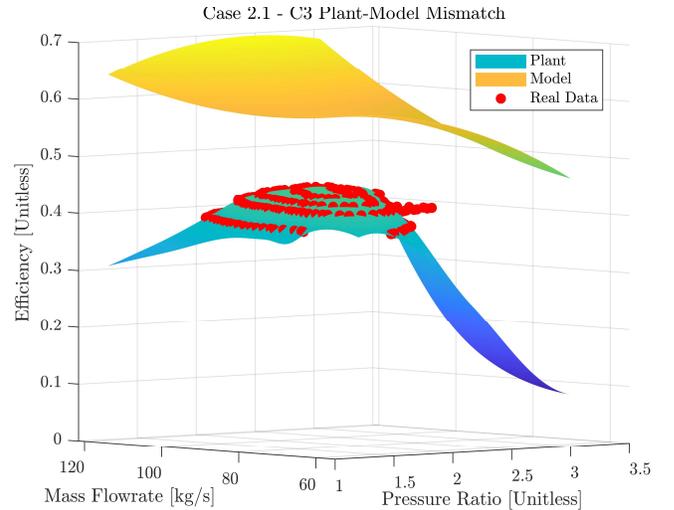

Figure 5. A mismatch between the model (orange surface) and the plant (blue surface) exists in this case. The mismatch was induced by the applied shift and by using only two points to fit the polynomial function from (7).

### D. Case 4

Case 4 simulates the other scenario described in section II.B where no prior model of the compressor map is assumed and instead this is learned directly from data by the GP as per (12) and (13). As a result, the model adaptation block now fits the GP directly to the efficiency surface. Similarly, the optimization block uses the fitted GP model of the compressor map directly.

V. RESULTS AND DISCUSSION

Fig. 6 shows the plots of the power consumption of the compressor station over three days of simulation for the different cases discussed in section IV. In each of the figures, the green solid line represents case 1, the best-case scenario with no mismatch. Similarly, the coloured solid lines represent cases 2.1, 2.2 and 2.3 which act as upper bounds to the problem while the dashed coloured lines demonstrate the performance of the GP error functions in Fig. 6.1, 6.2 and 6.3 while Fig. 6.4 exhibits the performance of case 4.

Fig. 6 shows that the GPs perform well in each of the cases where they are employed. The power consumption of these cases approaches that of case 1 within only a few set-point changes suggesting the GPs are able to learn from the data. As expected, as the degree of mismatch decreases from case 2.1 to case 2.3, the faster cases 3.1-3.3 approach the performance of case 1. This is expected because as demonstrated in Fig. 5, which exhibits the worst mismatch for case 2.1, the GP must learn an error function which overcomes this major mismatch. This will only begin to occur when a sufficient amount of data is collected from the plant such that the error function can be appropriately approximated by the GP. Therefore, for case 3.1 (red dashed line) shown in Fig. 6.1, there are early periods of operation where this tends to the worst-case performance of case 2.1 (red solid line). However, after only a few subsequent set-point changes, case 3.1 approaches the performance of case 1. This highlights the robustness and flexibility of using GPs in the *model-parameter adaptation* approach.

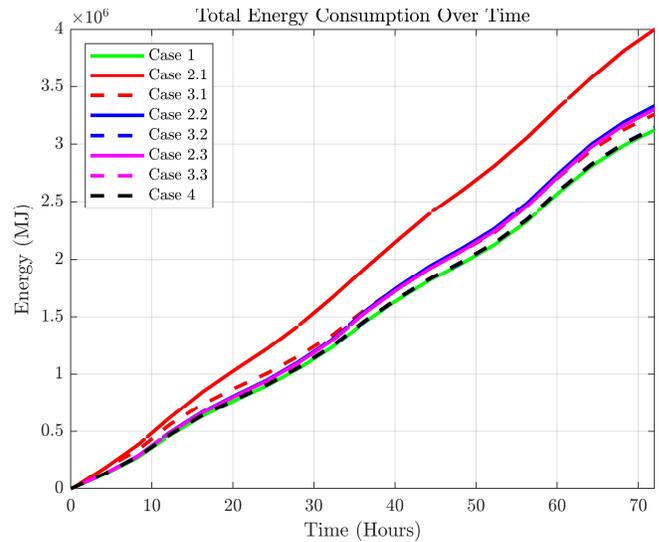

Figure 7. Total energy consumption of the compressor station for each of the cases defined. The performance of case 3.1 demonstrates the adaptive nature of GPs as they continually learn from data.

The adaptive nature of the GPs as they learn more about the system as the operation proceeds is best demonstrated with Fig. 7. Fig. 7 shows the total energy consumption of the compressor station over time for each of the different cases. In summary, it shows the cumulative effect of the GPs as they continually learn as new data becomes available. In particular, the cumulative improvement of case 3.1 exhibits the favourable properties of GPs which make them amenable for use in the *model-parameter adaptation* approach. Despite a major mismatch between the plants and models, the error functions represented by the GPs become a better fit to the real error functions as more data becomes available. Consequently, as shown in Fig. 7, the energy consumption of the whole station is driven towards case 1, the best-case scenario despite starting near the worst-case scenario of case 2.1. Alternatively, if the *model-parameter adaptation* approach were to be applied to a parametric model, one might expect the cumulative improvement to eventually plateau. This is because at some point, with an inadequate model structure, no updating of the parameters would minimize the remaining structural mismatch as discussed in [8] and [9]. However, as non-parametric models, GPs do not have this issue.

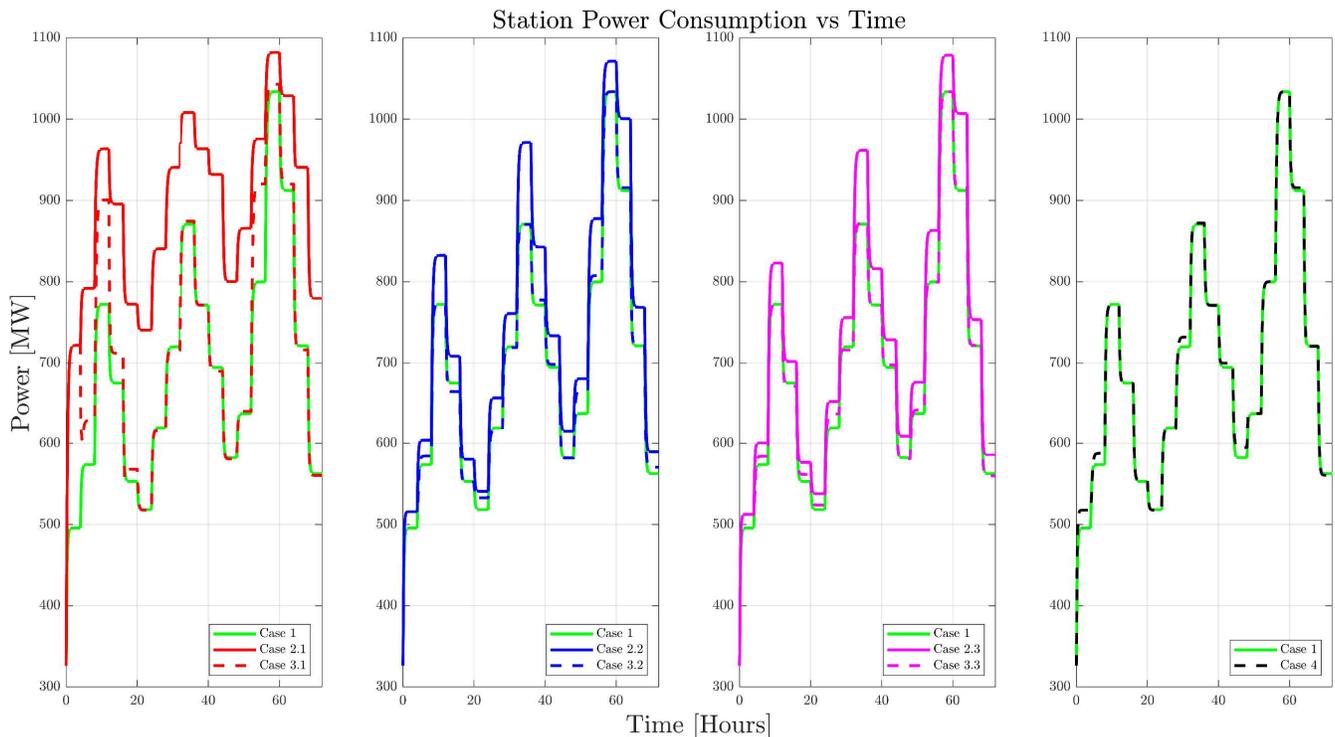

Figure 6. Power consumption of the compressor station over three days of simulation for the different cases discussed in section IV. In each of the figures, the green solid line represents case 1, the best-case scenario with no mismatch. Similarly, the coloured solid lines represent cases 2.1, 2.2 and 2.3 which act as upper bounds to the problem while the dashed coloured lines demonstrate the performance of the GP error functions in Fig. 6.1, 6.2 and 6.3 while Fig. 6.4 exhibits the performance of case 4.

The performance of case 4 is particularly insightful. Case 4 learns the compressor map directly from data and has no prior model as a starting point. The results of Fig. 6.4 and Fig. 7 once again show that the performance of case 4 approaches case 1 within a few set-point changes and in fact this occurs faster than case 3.1. This suggests that the utility of having a prior model is lost if the model used is not able to approximate plant behaviour to some extent. This is because, at the start of operation, when little to no observations have been made, the model of the compressor map when a GP error function is used is heavily reliant on the prior model. As a result, if the prior model is a poor representation of the plant, as is the case for the compressors in case 3.1, the model of the compressor map only converges onto the plant compressor map after a number of observations have been made. That is, only once evidence is collected to suggest that the prior model is a poor representation of the plant does the error function represented by the GP begin to correct this. This is clearly demonstrated by the cumulative improvement of case 3.1 in Fig. 7 and is in line with the Bayesian approach adopted by GPs as discussed in [10]. This is a conclusion supported by previous studies which attempt to incorporate prior knowledge into a hybrid modelling scheme [14] [15].

This concept is further accentuated in Fig. 8. Fig. 8 shows the evolution of the model of the compressor map for compressor 2 for case 3.1 as more data is observed over time. The red surface represents the prior polynomial model, while the green surface represents the plant compressor map. The adaptive model compressor map, consisting of the prior model and the error function, is represented by the blue surface. Finally, the operating points are represented in black. At the start of operation, shown in Fig. 8.1, as only few operating points (black) have been observed, the model compressor map (blue) locally approximates the plant compressor map (green) near these points. On the other hand, far away from these operating points, the model compressor map is heavily biased by the prior model (red) and so approximates this polynomial model instead. However, at the end of the simulation as shown by Fig. 8.2, where a significant amount of data has been observed, the model compressor map becomes a better global approximation to the plant compressor map but with some local differences biased by the prior model where no observations have been made. This demonstrates that despite the major mismatch initially, after data is collected, this mismatch is mostly corrected especially along the points of operation for the compressor. Additionally, it will be noticed that the operating points of the compressor lie on a curve spanning the compressor map. This is referred to as the system resistance curve and defines the relationship between pressure and flow within the system. For this reason, as only points along this curve are observed, the GPs must extrapolate outside of this range. Consequently, having a good prior model can ensure good global approximation of the plant compressor map while, more importantly for optimization, the GP error function can ensure good local approximation along the resistance curve. This is the main requirement of the model when solving the optimization problem formulated in (4) as the constraints defined in (4b) guarantee that the solution will lie on this curve. Therefore, having a good local approximation along the resistance curve, which the GP error function provides, ensures that an accurate solution to the optimization problem can be found. However, should the resistance curve change during operation, for example due to equipment changes, then the GPs would be able to adapt accordingly. In this case they would be able to learn a better global approximation of the true error function as more of the compressor map would be operated, further highlighting the flexibility of GPs.

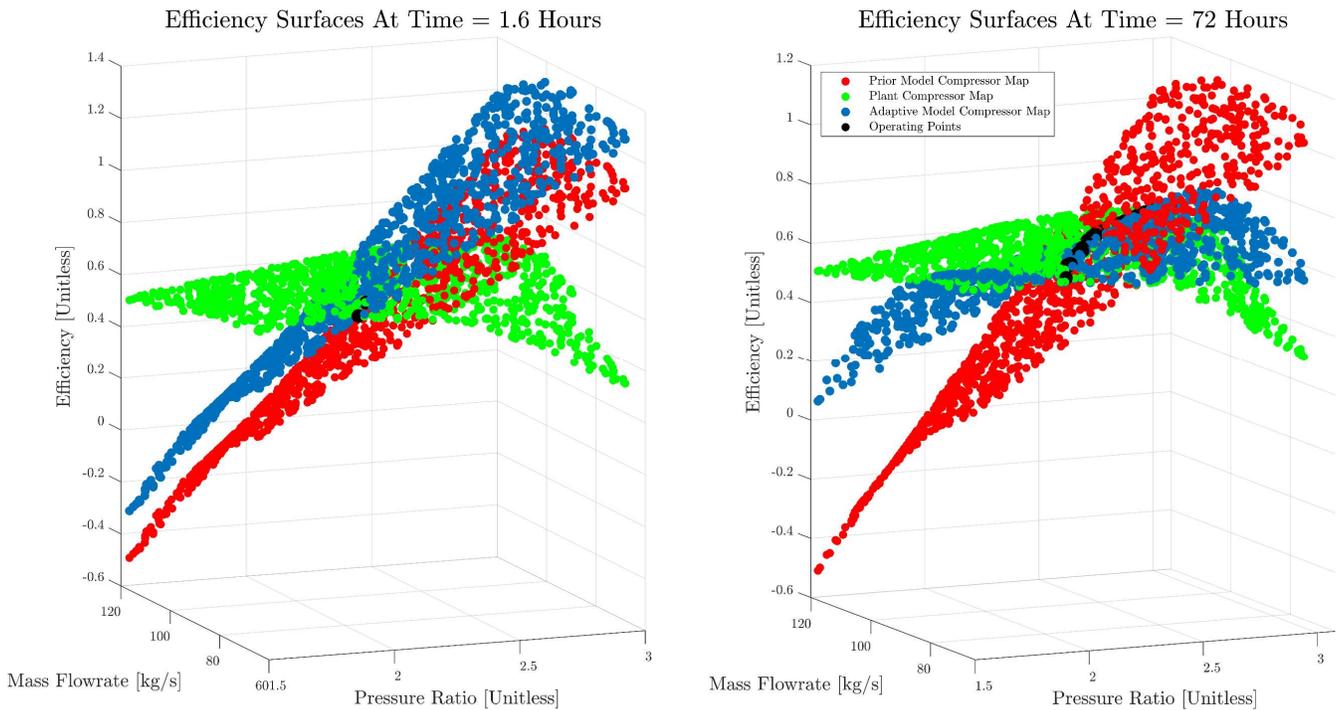

Figure 8. Evolution of the model compressor map (blue surface) for compressor 2 in case 3.1 over time as more data is measured from the system. The red surface represents the prior polynomial model, the green surface represents the plant compressor map and the model compressor map, consisting of the prior model and the error function, is represented by the blue surface. Finally, the operating points are represented in black.

Finally, in this paper, although we have managed to apply GPs to the problem of load sharing optimization successfully, as mentioned in section III.C, GPs may struggle to scale for larger problem sets. In our case, this is a relatively small-scale problem from the perspective of the dimensionality and amount of data. Even with more compressors, scalability would not be an issue as an additional compressor requires only an additional GP. Similarly, the Euclidean distance metric used in the simulation as described in section III.C ensures that the size of the dataset used to train the GPs does not become excessive. Therefore, one potential area of further investigation lies in the scalability of GPs to larger problems where datasets will likely grow significantly. It would be worthwhile as future work to investigate the application of GPs to large scale problems and in particular the use of approximation methods such as sparse GPs to deal with the problem of scalability [10] [16]. More specifically, the favourable properties of GPs could be exploited for large-scale real-time optimization problems if the issue of scalability can be resolved.

## VI. Conclusion

In this paper, we have used Gaussian Processes in the *model-parameter adaptation* approach to solve the load sharing optimization problem for compressor stations. In doing so, the results in this paper give strong evidence that GPs are robust in the face of varying degrees of model uncertainty and mismatch. In particular, due to their flexible nature as non-parametric models, we show that they would be more suited to the *model-parameter adaptation* approach than parametric models which may retain structural mismatch with the plant. This supports the conclusions made by [9]. Additionally, the results of case 4 also support a purely data-driven approach when an adequate prior model of the system does not exist. However, as demonstrated by the results of case 3.2 and particularly 3.3, where a prior model exists, for which there is relative confidence, then its use can be beneficial. In such a case, the GP plays the role of minimizing any structural mismatch which may exist between the prior model and the plant. This could be a useful area of exploitation for systems in which current models are sufficient but perhaps miss some important physics which are not fully understood and thus not being captured. Consequently, the GP can be used in a way to learn the missing model structure from data. Finally, although this paper focused on the application of GPs to the load sharing optimization problem applied to compressor stations, there is nothing to suggest that GPs could not be used for a more general class of problems found in the real-time optimization literature. Additionally, there are many other applications which rely on the modelling of complex performance curves for the control and optimization of machinery and processes. The approaches used in this paper could be utilized in such applications.